# Economic controls co-design of hybrid microgrids with tidal/PV generation and lithium ion/flow battery storage

Jonathan Cohen[1], Michael Kane[2,*], Alexia Marriott[3], Franklin Ollivierre III[4], and Krissy Govertsen[2]

1  Northeastern University, Department of Electrical and Computer Engineering; Boston, MA, USA
2  Northeastern University, Department of Civil and Environmental Engineering; Boston, MA, USA
3  Waltham High School; Waltham, MA, USA
4  Milton High School; Milton, MA, USA
*  Correspondence: mi.kane@northeastern.edu;
   400 Snell Engineering; 360 Huntington Ave.; Boston, MA 02115.

## ABSTRACT

Islanded microgrids powered by renewable energy require costly energy storage systems due to the uncontrollable generators. Energy storage needs are amplified when load and generation are misaligned on hourly, monthly, or seasonal timescales. Diversification of both loads and generation can smooth out such mismatches. The ideal type of battery to smooth out remaining generation deficits will depend on the duration(s) that energy is stored. This study presents a controls co-design approach to design an islanded microgrid, showing the benefit of hybridizing tidal and solar generation and hybridizing lithium-ion and flow battery energy storage. The optimization of the microgrid's levelized cost of energy is initially studied in grid-search slices to understand convexity and smoothness, then a particle swarm optimization is proposed and used to study the sensitivity of the hybrid system configuration to variations in component costs. The study highlights the benefits of controls co-design, the need to model premature battery failure, and the importance of using battery cost models that are applicable across orders of magnitude variations in energy storage durations. The results indicate that such a hybrid microgrid would currently produce energy at five times the cost of diesel generation, but flow battery innovations could bring this closer to only twice the cost while using 100% renewable energy.

### 1.1 KEYWORDS:

hybrid microgrids; optimization; renewable energy sources; tidal energy; solar energy; energy storage systems; lithium-ion batteries; vanadium redox flow batteries

## 1.2 HIGHLIGHTS

- Premature battery failure is an important aspect in microgrid design
- Flow battery costs vary significantly with energy to power ratio
- Controls codesign is the concurrent design of system size and controller parameters
- The LCOE of tidal-PV and LIB-VRFB microgrids is currently 4x diesel generator LCOE

# 2 INTRODUCTION

Many difficulties arise when integrating renewable energy sources (RES) into the grid due to their variability and unpredictable nature. Periods of peak production and consumption rarely align, making energy storage systems (ESS) necessary to balance differences between supply and demand [1],[2] and to ensure the energy supply remains stable and reliable [3],[4].

An ideal ESS has a long lifespan to minimize the cost of replacement, a high-power density to handle rapid power fluctuations, and a high energy density to smooth out variations in generation and load. However, a single energy storage technology is unlikely to meet all these requirements in an economical manner. This presents an opportunity for hybrid ESSs that utilize the best characteristics of different ESS chemistries [5].

Different batteries have different benefits. Vanadium redox flow batteries (VRFBs) have relatively low costs per energy stored, can easily be scaled up, do not undergo increased degradation due to deep discharge, and have a broader state of charge range than lithium-ion batteries (LIBs) [6], [7]. Estimations also show that VRFBs may be cheaper to produce than LIBs for long-term energy storage applications. However, this advantage is currently offset by their low production volume [8]. LIBs, which have a high power density, are also a promising energy storage option that is currently being produced at scale for grid and electric vehicle applications [9], [10]. However, high cycling rates, overcharge, and deep discharge each increase aging in LIBs [11], [12]. Flow batteries are primarily used for long term energy storage, while LIBs are used to deliver energy and quickly respond to demand. Hybridization can decrease operational stress and increase battery lifetime, thus reducing the levelized cost of energy (LCOE) delivered [13], [14].

Increasing the diversity of generation sources could increase the likelihood energy is produced when needed, reducing the need for battery capacity and cycling [15]. Solar photovoltaic (PV) arrays are highly modular and easily scalable [4]. However, electricity production using solar PV arrays fluctuates with changing weather conditions that are difficult to predict [16]. By contrast, tidal power, which uses energy from the ebb and flow of the tides to generate electricity, is as predictable as the moon's cycles and orders of magnitude less affected by local weather. However, the scaling of tidal systems can be significantly constrained by local hydrology.

Microgrids are electrical grids capable of producing and distributing power throughout a localized area. When islanded, they can do so without external control or energy [17]. Microgrids improve service quality and enable RES grid integration by lowering transmission losses and the time needed to fix outages [18]. These characteristics are especially valuable to island communities which traditionally rely on diesel power generation. Due to the high costs of importing fossil fuels, islands can handle the considerable expenses associated with renewables, energy storage, and first-generation microgrids [19].

While much of existing literature on hybrid-generation and hybrid-storage in microgrids covers the joint optimization of wind and solar, a relatively small number covers tidal energy or the use of VRFBs in hybrid ESSs. Furthermore, only a few studies account for use-based reductions in battery lifetime, resulting in overestimating battery revenues [20]. In [3], performance models for a PV-wind system with LIB storage are developed based on data representative of a location in Denmark. In [21], particle swarm optimization (PSO) is used to minimize the cost of energy of a wind/tidal/PV hybrid energy system. In [22], a novel expert fuzzy system-grey wolf optimization method is used to minimize operating costs and $CO_2$ emissions and maximize the efficiency of a microgrid consisting of a PV system, wind turbine, tidal turbine, and diesel generator using only LIB in the ESS. In [15], an integrated energy system with combined heat and power generation, PV, and battery energy storage is optimized while taking into account battery lifetime loss using a simple total power throughput degradation model. In [5], a hybrid wind/PV and battery/supercapacitor microgrid system is optimized to minimize costs and greenhouse gas emissions and improve reliability without accounting for abnormal battery degradation. In [23], the proprietary HOMER software is used to optimize a PV/wind hybrid power generation system. In [24], an operational planning strategy is defined for an islanded microgrid containing tidal, PV, and fuel cell generators with only thermal storage (i.e., storing the heating the fuel cells). In [25], a short-term scheduling algorithm is presented for a tidal-powered microgrid with LIB storage with a lifetime modeled as a nonlinear function of depth of discharge.

This manuscript builds on this body of literature tying together four main contributions. (1) This is the first paper of its kind to study the economic benefits of combining solar PV with tidal generation and LIB with VRFBs. (2) A filter-based control algorithm is proposed to determine the appropriate (dis)charge out/into each battery type. A controls co-design approach [26] is used to simultaneously optimize the control algorithms' parameters and the physical RES and ESS sizes. (3) Battery life is modeled not only using a maximum lifetime in years but also used cycle life. Furthermore, different cycle life models are used for LIB and flow batteries. (4) Lastly, the companion code released with this manuscript provides a flexible simulation environment for future research on hybrid microgrids.

The rest of this manuscript is organized into sections on Materials and Methods (§2), Results (§3), Discussion (§4), and Conclusion (§5). The Materials and Methods section presents the model architectures and parameters considered for the RES and ESS subsystems, the overall microgrid model, optimization methods used, and the proposed parametric studies of cost and performance. The Results section presents the LCOE of each simulation and the contribution of each subsystem to the LCOE; time series plots of the energy flows for key simulations, and the results of the parametric studies. The Discussion section draws insights from the parametric study results to discuss the opportunities for such hybrid microgrids under various cost scenarios. The Conclusion summarizes the manuscript and presents challenges that should be addressed in the future to help realize such hybrid microgrids.

# 3 MATERIALS AND METHODS

This manuscript considers a study of the microgrid system shown in Figure 1 providing electricity to a large island community using hybrid RES from tidal and solar power and a hybrid ESS with LIB and VRFB modules. A microgrid battery controller is designed to allocate excess generation to charge the batteries and meet any deficit by discharging the batteries in a way that best leverages the unique physics and economics of each battery type. This work is entirely simulation-based, developed with object-oriented programming in MATLAB [27]. Each simulation is 1-year long and considers each discrete hour in the year. The model contains demand, RES, ESS, and control subsystem models, each with energy balance, economic, and component degradation models.

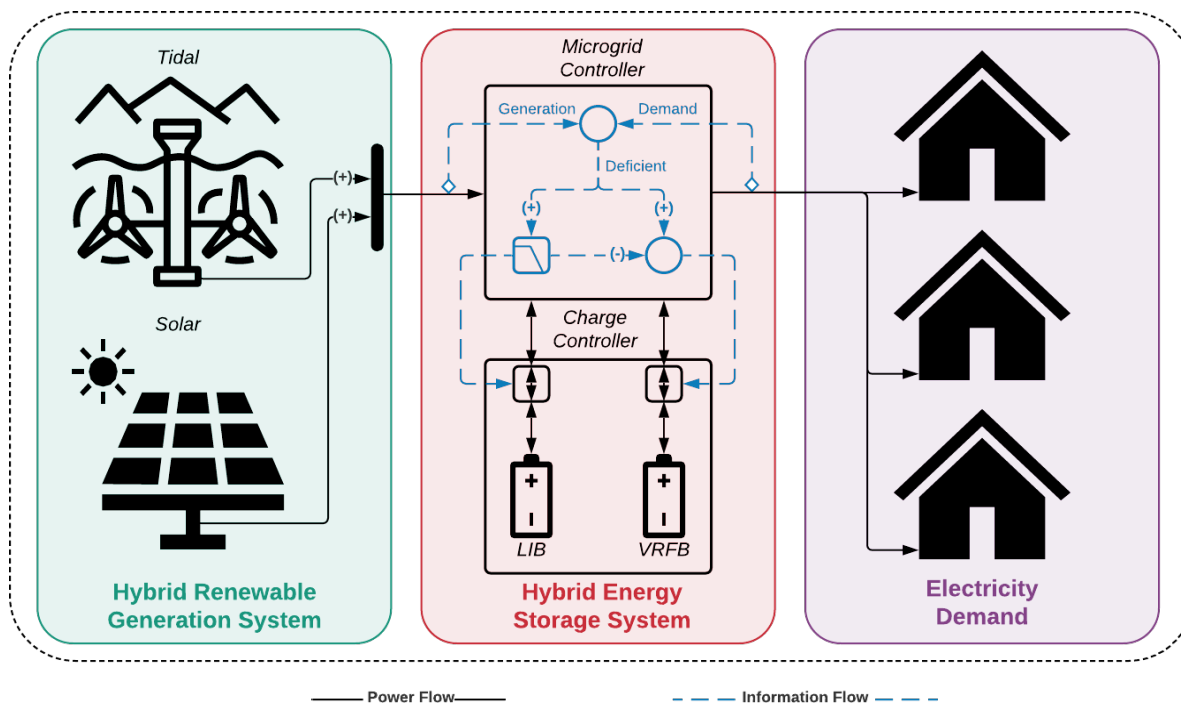

*Figure 1: Schematic overview of the hybrid microgrid using solar and tidal RES and LIB and VRFB ESS, with battery (dis)charging managed by the microgrid controller to meet the island's electricity demands.*

## RENEWABLE GENERATOR MODELS

### 3.1 SOLAR PV

The hourly energy generation of the solar PV system was modeled using the National Renewable Energy Laboratory's PVWatts Calculator [28] for the ZIP code 02807 (i.e., New Shoreham, RI, USA on Block Island) with a 1 kW DC system size with the default 14.08% system losses, 96% inverter efficiency, standard modules (i.e., ~15% nominal efficiency) in a fixed open rack at 20° tilt and 180° azimuth, resulting in a 15.9% capacity factor. The design parameter for the solar PV system is how much the 1 kW system should be scaled to meet demand. The installed cost per rated power is assumed to be $1,060/kW in the baseline scenario with a system lifetime of 30 years [29]. Operations & maintenance costs are neglected.

## 3.2 Tidal

The tidal energy system's hourly energy generation is assumed to be proportional to the tidal flows with the rated power output produced at peak flow. The tidal flow is modeled as the product of two sine waves with periods of a lunar day and a lunar month.

A lunar day is the amount of time required for a specific location on Earth to rotate from a point beneath the moon back to this original spot. The moon revolves around the Earth in the same direction as Earth's rotation on its axis, so due to Earth's additional time to reach the same location beneath the moon, lunar days are 50 minutes longer than solar days. Every lunar day, 2 high tides and 2 low tides occur [30]. This is modeled using a 6.2-hour wavelength sine wave with a minimum of zero and a maximum of 1.0, shown below where $t$ is the hour of the year.

$$\frac{\sin\left(\frac{t}{6.2 * 2\pi}\right) + 1}{2} \qquad (1)$$

The sun also causes smaller tides to occur. When the Earth, sun, and moon line up, the lunar and solar tides reinforce each other during full and new moons. Unusually small tides known as neap tides occur when the solar and lunar tides act against each other, while unusually large tides, known as spring tides, occur when solar and lunar tides reinforce each other. These high and low tides occur approximately every two weeks [31]. This effect is modeled using a 360-hour wavelength sine wave.

$$\frac{\sin\left(\frac{t}{360 * 2\pi}\right) + 1}{2} \qquad (2)$$

The design parameter for the tidal RES is the generator's rated power. The installed cost per rated power is assumed to be $4,300/kW in the baseline scenario with a system lifetime of 20 years [32].

## 3.3 Battery models

Two types of batteries are considered for this study: conventional LIBs and VRFBs. In general, the battery model is founded on an energy balance rule (i.e., stocks and flows) at each hour of the year. In an hour where the island's electric demand is greater than the supply from the RES, the deficit must come from the combined ESS, and when supply is greater than demand, the excess is stored in the combined ESS. Each battery has a maximum lifetime, in years, and a maximum number of charge/discharge cycles, where a 'cycle' is defined differently for the different battery types. Based on the supply-demand deficit/surplus at each hour, the battery controller (described in §2.3.1) allocates energy to (dis)charge each battery.

The cost of each battery is specified as the sum of the cost per energy storage capacity in $/kWh (e.g., energy capacity capital cost and construction and commissioning) and the cost per rated power in $/kW (e.g., power conversion system and balance of plant) [33]. Both types of batteries are assumed to have a round-trip efficiency of nearly 100%.

### 3.3.1 Lithium-Ion Battery (LIB)

Due to their relatively high energy capacity costs and relatively low rated power costs, LIBs are best suited for short, high power applications. In this model, the LIB ESS has a baseline energy capacity cost of $285/kWh (i.e., $189/kWh for the capital cost of the energy capacity of the battery itself plus $96/kWh construction and commissioning cost) and a rated power cost of $306/kW (i.e., $211/kW for the power conversion system plus $95/kW for the balance of plant). These costs are based on their 2025 estimate for utility-scale operations with an energy over power ratio of 4.0 [33]. The total energy capacity costs are assumed to scale linearly (i.e., at a fixed $/kWh) due to LIB packs' modularity. At the scales considered (e.g., MWhs of storage), the module level rated-power costs are assumed to be independent and negligible with respect to module-level energy capacity costs [34].

The LIBs have a maximum lifetime of 10 years, a cycle life of 3,500 cycles [33], and are replaced when the first of these limits is reached. A simple discharged-based model is used to estimate cycle life depletion. For example, a single cycle of a LIB with a 1 kWh capacity is used if fully charged then fully drained; or twice repeatedly (partially) charged and then 0.5 kWh of energy discharged; or four times repeated (partially) charged and 0.25 kWh of energy discharged.

### 3.3.2 Vanadium Redox Flow Battery (VRFB)

Due to their relatively low energy capacity costs and relatively high rated power costs, VRFBs are well suited for long term storage. VRFBs have two significant cost drivers at the module level: the aqueous electrolytes and the membrane and electrodes. Combined, these costs scale inverse-exponentially with respect to the energy over power (E/P) ratio according to (3). This equation was fit to the data for VRFB capital cost in [35], reproduced in Figure 2. At very large E/P (i.e., large durations of storage/discharge), the cost is governed by the electrolyte costs; while at small E/P (i.e., large power output) the cost is controlled by the membrane and electrode costs.

$$C_{module_{VRFB}} = (7.004e+04)e^{\left(\frac{0.004021}{E/P}\right)} - (6.9837e+04) \tag{3}$$

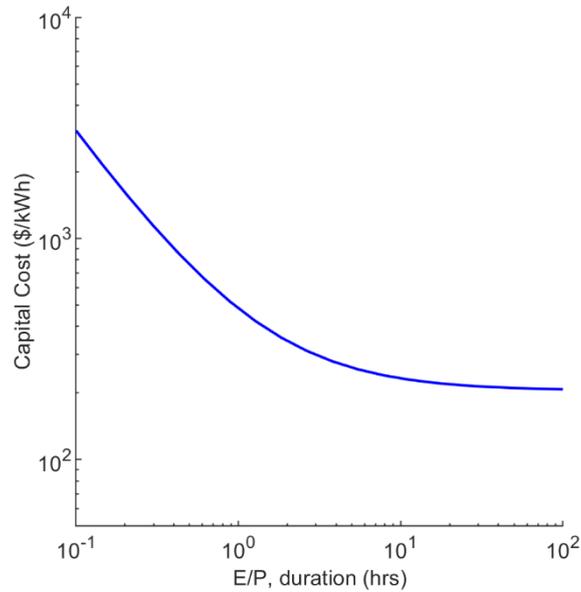

*Figure 2: Effect of E/P ratio on VRFB module capital costs.*

The VRFB module capital costs are only part of the total VRFB ESS costs. Additional costs, including construction and commissioning, $650/kWh; power conversion system, $211/kW, and balance of plant, $95/kW, are considered [33]. The cost estimates in [33] assume E/P = 4.0; at this size, they estimate a energy-capacity capital cost of $393/kWh, while the more general model from [35] provides an estimate of $278/kWh. The sensitivity of this cost parameter will be studied later in this manuscript.

The VRFBs have a maximum lifetime of 15 years, a cycle life of 10,000 cycles [33], and are replaced when the first of these limits is reached. A cycle is depleted each time the battery switches from charge to discharge mode, due to the assumed degradation of the membrane. The higher cost of rated power output and the desire to limit (dis)charge mode switching make VRFBs best suited for long-duration energy storage.

## 3.4 Microgrid model

This study considers a microgrid designed to provide electricity to the 429 households [36] on Block Island. Due to the lack of a public load profile for this community, the load is assumed to follow the same shape as the hourly loads from January 1, 2019 to December 31, 2020 on the wholesale electric grid at the nearest load zone—i.e., ISO New England load zone 4005.Z.RHODEISLAND. This hourly load profile is then scaled to a total yearly energy consumption of 4.57 GWh, i.e., the energy use of the 429 households using the US average annual household electric energy consumption of 10.65 MWh [37]. The microgrid consists of a microgrid controller, the tidal and/or solar PV RES, and the LIB and/or VRFB ESS.

### 3.4.1 Microgrid Controller

The microgrid controller must ensure that the electricity demand is met at each hour of the year. The demand is assumed to be uncontrollable, and while the renewable generators can curtail energy, they cannot produce more than is available from the sun and tides at that hour. This leaves the microgrid controller responsible for splitting the charging (discharge) power surplus (deficit, respectively) between the demand and generation at each hour. An expensive backup generator is only available as a measure of last resort. The system should be designed such that this backup generation is not needed.

Considering the high-power benefits of LIBs and the long-duration benefits of VRFBs, the battery power-flow controller implements a low-pass filter [5] on the deficit $P_{deficit}$ between demand and generation at each hour $t$, allocating the high-frequency component $P_{LIB}$ to the LIBs and the low-frequency component $P_{VRFB}$ to the VRFBs. Following this sign convention, $P_{deficit}$ is negative when generation is less than demand; and $P_{LIB}$ and $P_{VRFB}$ are positive when the battery is discharging. The filter is implemented as a discrete-time (with one-hour time steps, $k$) causal moving-average filter, where the span of the moving average $K_{control}$ (in hours) is a design variable.

$$P_{VRFB}(t) = \frac{1}{K_{control}} \sum_{k=1}^{K_{control}} P_{deficit}(t - k - 1) \tag{4}$$

$$P_{LIB}(t) = P_{deficit}(t) - P_{VRFB}(t) \tag{5}$$

### 3.4.2 Design Objective

When designing the microgrid, the objective is to reduce the levelized cost of energy (LCOE) delivered by the microgrid. Since RES and ESS's operational costs are orders of magnitude less than those of traditional generation, only capital costs are considered in this analysis. To simplify the analysis and to avoid the calculation of salvage values, future costs are not discounted. Therefore, the LCOE is calculated according to (5) as the sum of capital cost $C_n$ of each component $n$ divided by each component's realized lifespan $T_{max_n}$ in years (e.g., if batteries are heavily cycled, they may need replacement before the maximum lifespan) divided by the total energy delivered per year ($E_{deliv} = 4.57\ GWh$) resulting in an LCOE in units of $/MWh delivered. To facilitate analysis and understanding of each microgrid component's impact, the LCOE is decomposed into the contributing LCOE of each component by dividing that component's capital cost by its realized lifetime divided by the total energy delivered by the microgrid.

$$LCOE = \frac{\sum_n^N (C_n / T_{max_n})}{E_{deliv}} \tag{6}$$

### 3.4.3 Design Problem

The three independent variables in the design problem are the rated power of the tidal generator, $[P]_{tidal}$; the rated power of the solar PV system, $[P]_{solar}$; and the span of the moving average filter in the power-flow controller, $K_{control}$. This leads to the following optimization problem. The lower limits of each of these variables are zero. While there may not be true upper limits on these variables, if a constrained optimization approach is used, the upper limits should be set high enough that they are not reached.

$$\min_{\substack{[P]_{tidal} \\ [P]_{solar} \\ K_{control}}} LCOE \tag{7}$$

The battery energy storage capacity and rated power are dependent variables, calculated as follows. With the hourly demand $P_{deliv}$ over the year known, and the rated power of the generators selected, the deficit $P_{deficit}$ is easily calculated at each hour. The battery controller then calculates the discharge power ($P_{LIB}$ and $P_{VRFB}$) of each battery (negative for charging) according to (4) and (5). The necessary rated power of the LIB battery $[P]_{LIB}$ is then calculated according to (7) (similar for VRFB) as the maximum absolute value of the observed (dis)charge power. Each battery's charge is then calculated according to (8) to ensure that the charge is always positive. The energy storage capacity $[E]_{LIB}$ is then calculated according to (9) to ensure the capacity is never exceeded (similar for VRFB). The shift means that the battery likely starts with an initial and final charge between empty and full. If the final charge is less than the initial charge, the difference must be economically made up from expensive backup generation.

$$[P]_{LIB} = \max_{t}|P_{LIB}(t)| \tag{8}$$

$$E_{LIB}(t) = \int_0^t P_{LIB}(t)dt + \min_{t \in [0,8759]} \left( \int_0^t P_{LIB}(\tau)d\tau \right) \tag{9}$$

$$[E]_{LIB} = \max_{t}(E_{LIB}(t)) \tag{10}$$

From these data, the cycles and realized lifetime of the batteries can be calculated, as well as the capital cost of the RES and ESS.

### 3.4.4 Grid search

An exhaustive grid search of any two independent variables shown as a 2-axis contour map of the LCOE can provide insights into the optimization problem's convexity, smoothness, and general shape. Four 'slices' of this three-dimensional optimization problem are selected for the grid search: (1) vary the tidal and PV RES rated powers using only the LIB; (2) vary the tidal and PV RES rated power using only the flow battery; (3) vary the rated power of the tidal RES and battery controller filter span, without the PV RES; and (4) vary the rated power of the PV RES and battery controller filter span, without the tidal RES.

Due to the large range of these variables under consideration (e.g., the solar rated power may vary from zero to 5MW), a log spacing is used to define the grid. The lowest LCOE on the grid is then selected as a starting point for an interior-point constrained local optimization [38] to further improve the lowest point's accuracy, which may lie between grid lines.

### 3.4.5 Particle Swarm Optimization (PSO)

An exhaustive search of the full three-dimensional space would be computationally intractable. Instead, a PSO method [39] is used to identify the system configuration ($[P]_{tidal}, [P]_{solar}, K_{control}$) with the lowest LCOE. A two-stage optimization approach is utilized, wherein the optimal point returned by the PSO initialized a secondary interior-point constrained local optimization [38] to refine the optimal configuration estimation further. As with the grid search, the independent variables are projected onto a logarithmic space. The swarm size is the primary meta-parameter of the optimization and should be selected as large as possible until computational tractability limits are reached or negligible improvements are realized.

### 3.4.6 Sensitivity Analysis

The system cost established above are estimates based on the latest literature. However, RES and ESS's cost has been decreasing exponentially in recent history, and new technology developments and business practices can yield significant changes in market prices. A sensitivity analysis will provide insight into how such price fluctuations may affect fundamental system architecture. Specifically, the study separately considers four separate cost variations: (1) the energy storage capacity costs of the LIB module; (2) the dielectric, membrane, and electrode costs of the VRFB module; (3) the cost per rated power of the solar PV; and (4) the cost per rated power of the tidal generator. The power conversion and balance of system costs are not varied in the sensitivity analysis as they are not expected to fluctuate as greatly.

In the four separate sensitivity analyses, the cost is varied from 1/10th the baseline cost to double the baseline cost in twenty equally spaced steps. At each step change in component cost, a PSO is conducted to determine the system configuration ($[P]_{tidal}, [P]_{solar}, K_{control}$) with the lowest LCOE. Plotting the components LCOEs in a stacked area chart will then show continuous and step changes and optimal system configuration as component costs change.

# 4 RESULTS

A simulation environment was developed in MATLAB for this work [40], with wrappers for the grid search, PSO, and sensitivity analysis. Using an object-oriented program approach yielded a software environment where microgrid components can be easily swapped out, added, and have their parameters modified. First, the grid search results are shown to provide an understanding of the optimization surface. The PSO algorithm is then tuned, yielding an efficient swarm size that is highly likely to identify global optimums in the search domain. A system is then designed with the PSO using the baseline parameters established above, producing insights into the key contributors of the LCOE and the time-domain response of the microgrid. Finally, the sensitivity of LCOE and optimal system configuration with respect to variations in component costs is presented.

## 4.1 Grid Search Results

All four grid searches (Figure 3) show the value of the hybrid microgrid, both hybrid RES and hybrid ESS. This is apparent in the lowest LCOE value, always balancing between solar and tidal rated power and a battery controller that utilizes both LIB and VRFB ESSs. The whitespace in the figure represents LCOEs greater than an enormous $100/kWh and therefore not relevant to show their surface profile. The optimal point (shown as a red dot) is at the base of this steep cliff in all four cases. Inside the boundary (down and to the left, primarily white space) is a microgrid that fails to generate all its own energy and/or the generators and batteries fail to deliver power at times resulting in the use of expensive backup generation. Outside the boundary (to the right and/or up), the costs increase steadily as the system becomes oversized, curtailing energy and/or having unused battery capacity. The battery filter span for a system with only solar RES is 13.7 hours and 64 hours for a tidal generator.

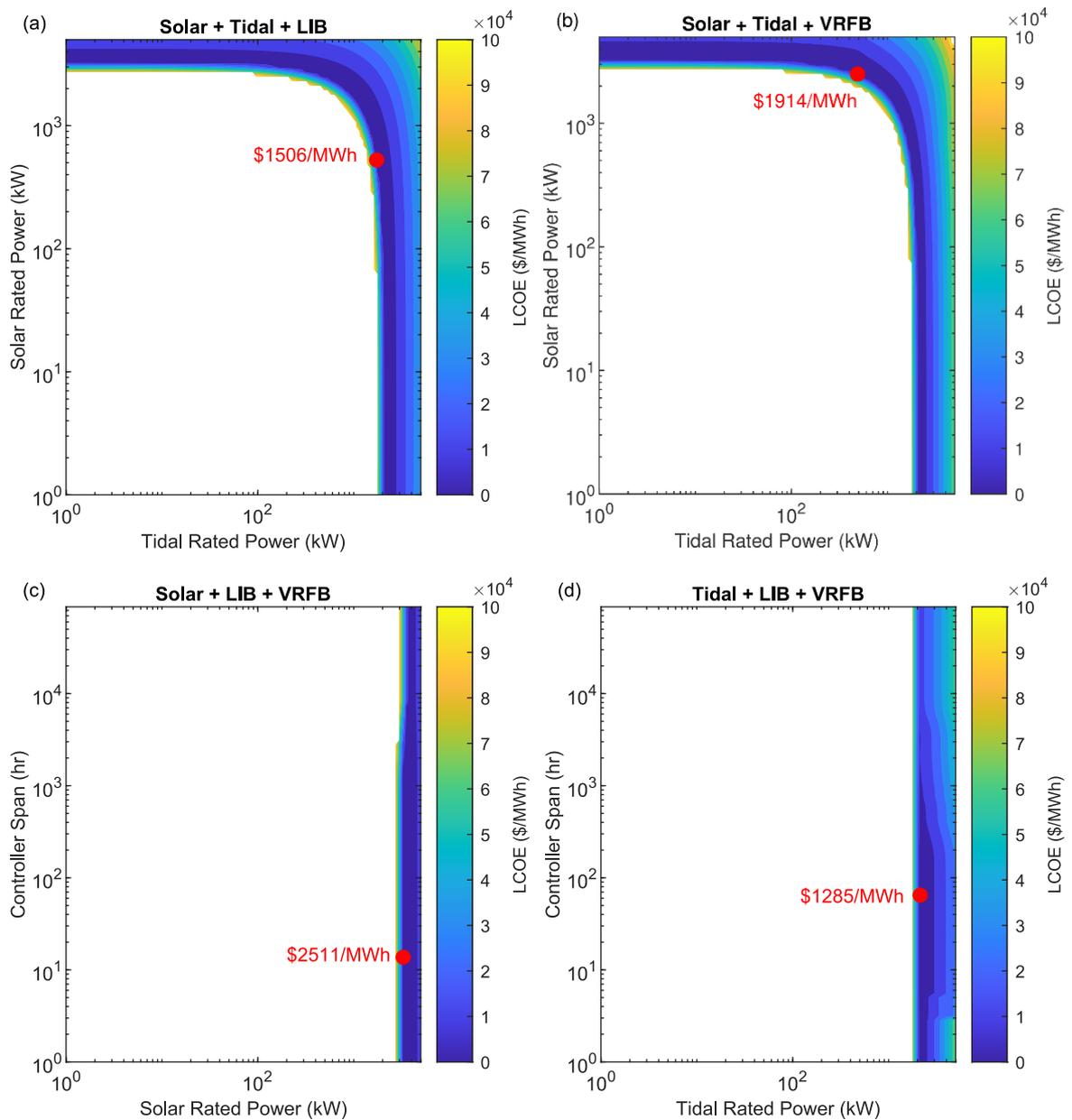

*Figure 3: Grid search results plotted as LCOE contours.*

*The red dot shows the lowest LCOE on the surface. Points on the grid with an LCOE greater than $100/kWh are not shown. (a) and (b) show the impact of solar RES rated power and tidal RES rated power on the LCOE of a system with only an LIB ESS and VRFB ESS, respectively. (c) and (d) show the impact on LCOE of the span of the battery controller moving average filter and the rated power of the solar and tidal RES, respectively.*

## 4.2 Selecting the PSO Swarm Size

The PSO algorithm's default swarm size is the minimum of 100 or 10 times the number of independent variables (i.e., 3) leading to a default swarm of 100. Rerunning the PSO with swarm sizes of 100, 266, 708, 1884, and 5012 led to less than a 0.1% variation in the solutions. Running on the PSO algorithm with parallelization on Northeastern University Discovery super-computing cluster with 65 nodes and 64GB of RAM, the computation time was reduced to ~40 seconds when a swarm of 200 was used. As such, a swarm of 200 was used for all the results shown in this manuscript.

### 4.3 BASELINE HYBRID MICROGRID

Using the baseline costs and parameters established above, the PSO algorithm reduced the LCOE to $1,186/MWh by selecting a 1.7MW rated-power tidal RES, a 0.5MW rated-power solar RES, and a battery controller moving average filter with a 15-hour span. The total cost is $83M. The microgrid served the full demand for energy of 5 GWh with the combined RES generating a peak power of 85 kW. The 1.7MW tidal RES cost $7.5M (at $4.3/MW) and produced 3.8 GWh/year, thus a 25% capacity factor. The 0.5MW solar RES cost $0.6M (at $1.1/MW) and produced 0.7 GWh/year, thus a 16% capacity factor. The resulting LIB ESS has an energy storage capacity of 3 MWh ($0.7M at $285/kWh) and a rated power of 1.0MW ($0.3M at $306/kW), thus a E/P ratio of 2.6 hours. The cycling reduced the realized lifespan from a maximum of 10 years to 7.9 years. The resulting VRFB ESS has an energy storage capacity of 225 MWh ($73.1M at $325/kWh) and a rated power of 0.8MW ($0.4M at $503/kW), thus a E/P ratio of 296 hours. The cycling was not enough to reduce the lifespan of the VRFB below the maximum of 15 years. The energy storage costs of the VRFB are the primary contribution to the $1,186/MWh LCOE as shown in Figure 4.

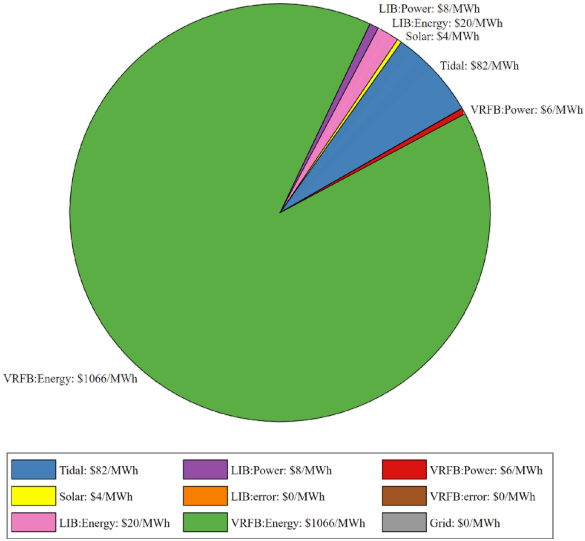

*Figure 4: The breakdown of each microgrid component's contribution to the total system LCOE.*

Figure 5 shows the demand, generation, and energy stored at each hour of the simulated year. In Figure 5(a) the monthly variations in the tidal power generation are seen, significantly exceeding the demand at times. Further, the daily fluctuations in solar generation are generally less than the demand. The curtailments (i.e., excess RES generation not used by the demand or ESS) and it's negative, the amount of power required from ancillary sources (e.g., diesel generation) is strictly and significantly less than 1.0mW. Figure 5(b) shows the energy stored by the VRFB is two orders of magnitude higher than that stored in the VRFB. Yet, the figure hides the significant power (i.e., 1MW peak) provided by the LIB.

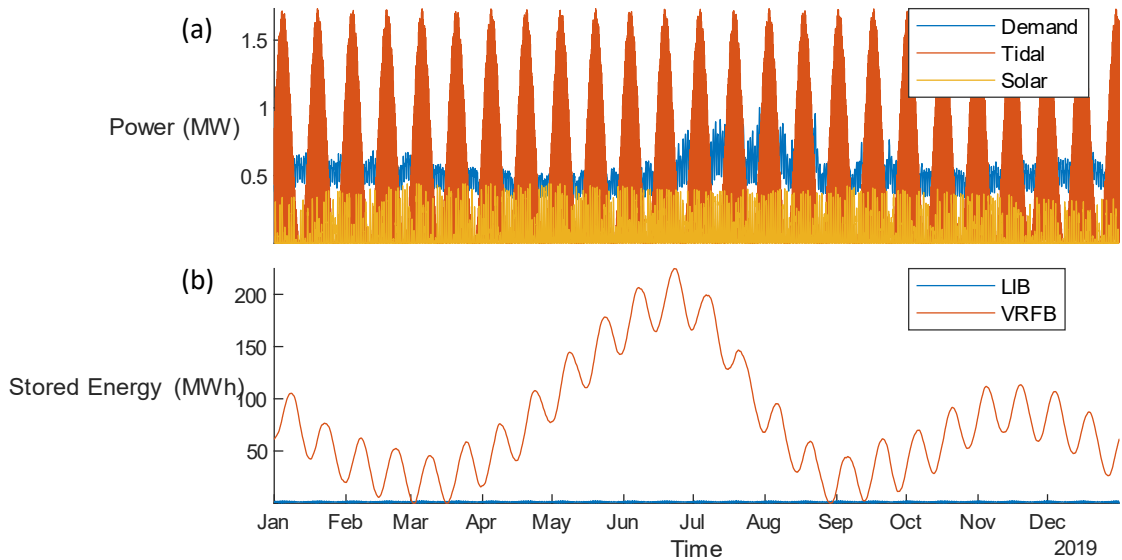

*Figure 5: The simulated generation, demand, curtailments, and energy stored each hour of the year for the optimal baseline-cost microgrid with tidal and solar RES and LIB and VRFB ESS.*

### 4.4 COST SENSITIVITY STUDY

Figure 6 shows the sensitivity of optimal system configuration to variations in the cost of system components. In Figure 6, very low LIB costs are shown to lead to significantly greater LIB energy storage. When the baseline cost is multiplied by 1.8, a large increase in LCOE is observed due to a large increase in the energy storage requirements. In general, the VRFB is the most significant component of the LCOE, thus the linear varying left side of Figure 6(b) is expected: the total LCOE would increase significantly as the VRFB costs are increased. As the VRFB costs are increased ~1.5x a step change in the use of LIB is observed to replace the VRFB storage, and at a 2x increase in VRFB the LIB becomes the most significant source of energy storage. The solar cost is varied in Figure 6(c); however, due to the relatively small contribution of the solar system to the total LCOE, even the large variations in solar costs lead to only linearly increasing system costs with no substantive changes in configuration. The tidal cost increases also lead to linearly increasing LCOE in Figure 6(d) without significant configuration changes. However, since the tidal generators provide most of the system's energy, the tidal energy cost increases lead to a more significant increase in total LCOE.

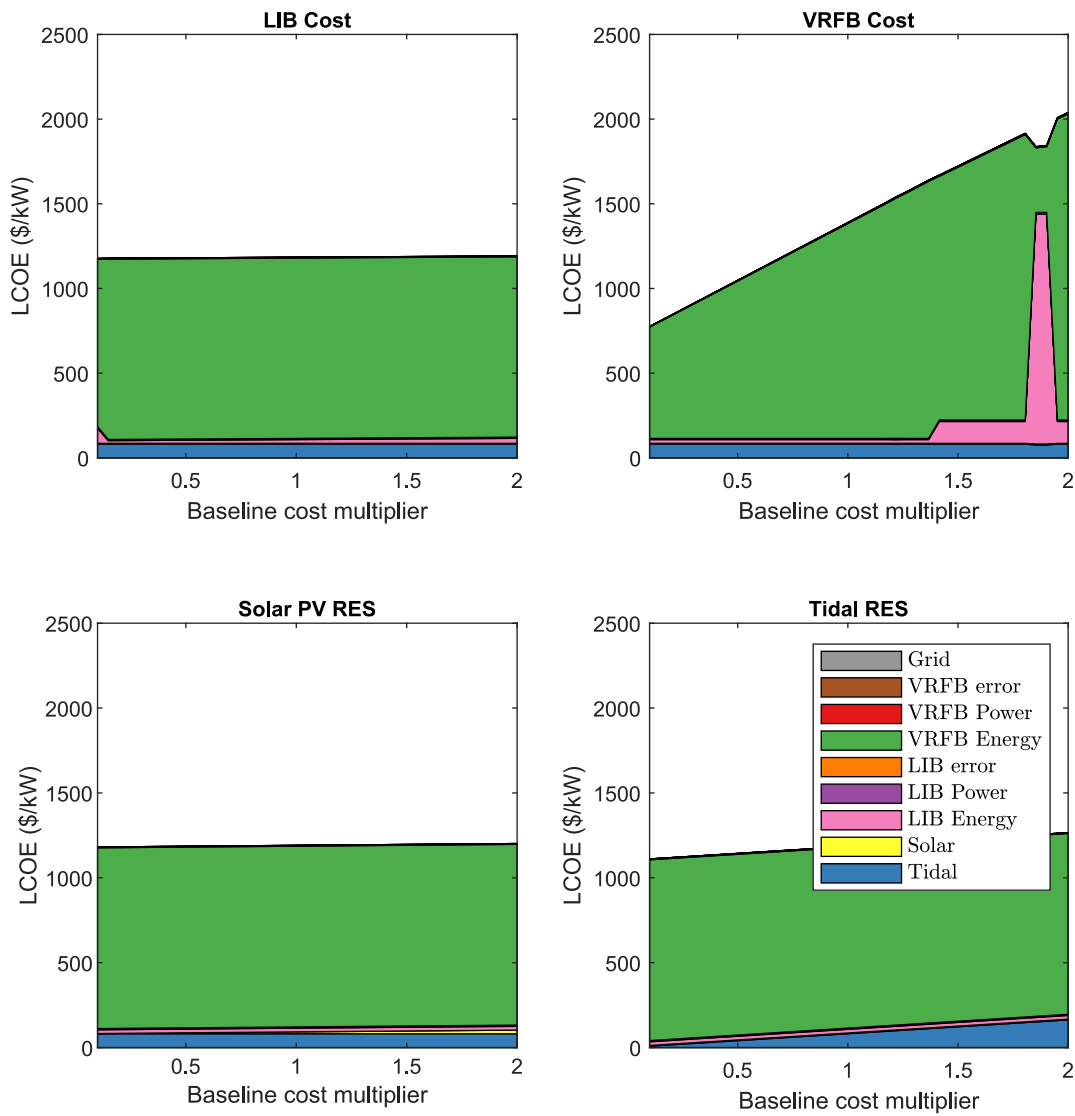

*Figure 6: Results of the cost parametric study where each component cost varies from 1/10th the baseline to 2 times the baseline. (a) LIB energy capacity cost variation. (b) VRFB module cost variation coming primarily from the cost of the electrolyte, membrane, and electrodes. (c) Solar PV RES rated power cost variation. (d) Tidal RES rated power cost variation.*

## 5 Discussion

The LCOE of the optimized baseline system was $1,191/MWh. This is significantly higher than typical wholesale electricity prices in the US of $20-$50/MWh [41]; typical retail prices in the US of $105/MWh [42]; and the LCOE of diesel reciprocating engines of $187-$319/MWh [43]. The majority of this high LCOE is due to the high cost of long-duration energy storage, even when using lower cost VRFBs. If the capital cost of VRFB modules can be reduced an order of magnitude through technological, economic, and business practice learning curves, the hybrid microgrid LCOE could be reduced to a more competitive $772/MWh. Unlike grid-tied electricity and diesel generators, the proposed hybrid microgrid is 100% renewable.

The utilization of both types of batteries highlights microgrids' value with hybrid ESSs and the importance of the microgrid's battery controller. When only a solar RES is considered (as in the grid search), the optimized battery controller had a moving average filter span of 13 hours, while when only considering tidal RES, the optimized span was 65 hours. As such, the solar microgrid utilizes the LIB to smooth out hourly variations in solar energy, and the VRFBs are used to provide energy during the dark nighttime hours and longer duration (e.g., seasonal) energy needs. On the other hand, the tidal microgrid uses the LIB to smooth out the daily tide variations and reserves the VRFB for the monthly tidal variations and seasonal load variations. Especially for microgrids with significant tidal generation, the microgrid battery control algorithm should be tuned simultaneously to optimize the tidal rated power since leveraging the wrong battery could significantly increase system costs and LCOE.

This study introduced two innovations missing in many previous studies of hybrid microgrids, battery lifecycle modeling and VRFB nonlinear cost modeling, both of which proved significant. The LIB and VRFB had maximum lifetimes of 10 and 15 years, respectively. However, excessive cycling drove premature replacement of the LIB in the baseline scenario, where it needed to be replaced in just 7.9 years. Most battery cost models cited in microgrid literature are specified only for a given E/P ratio, often around 4 hours [33]. This would likely be sufficient for LIBs since the E/P was 2.6 hours in the baseline scenario, and LIB storage capacity are easily scaled. However, according to Figure 2 a VRFB with an E/P of 4.0 hours would have a module cost around $274/kWh while the E/P of 296 hours used in the baseline has a module cost of $206/kWh. This is because the lower rated power results in smaller electrodes and membranes, reducing the battery cost.

The most economical system configurations were identified using the PSO global optimization algorithm. When implementing optimizations, it is helpful to understand the convexity and general smoothness of the optimization hypersurface. The grid search shows that, at least from 2D slices shown of the 3D optimization hypersurface, the optimization appears generally convex. In practice, a more computationally efficient gradient descent algorithm may be more desirable than the computationally costly PSO algorithm. However, this assumes that the initial point for the gradient descent algorithm is feasible and yields an acceptable LCOE and does not consider the possibility of not strictly convex portions of the surface. Furthermore, the cost sensitivity shows that the cost and system configuration is nonconvex and nonlinear with respect to component costs.

# 6  CONCLUSIONS

Microgrids using hybrid tidal and solar RES and hybrid LIB and VRFB ESS can provide economical energy to remote communities, provided the cost of VRFBs is significantly reduced as the technology and market matures. When studying and designing such systems, it is important to consider the realized life cycles of batteries due to excessive cycling; nonlinear relationships between cost, rated power, and energy storage capacity for VRFBs; and the simultaneous design of the physical system and controller parameters. This study contributes to this field by applying a simple lifecycle model for LIB and VRFBs; a VRFB cost model that accounts for the relationship between battery energy capacity and rated power; and implements controls co-design (CCD). To realize the promises of hybrid microgrids, future research should validate the approach with higher temporal fidelity; study the reliability and robustness of the design to temporal variations in RES generation and load; consider the time-value of money in the LCOE calculation; and implement demand response to minimize the need for energy storage.

# 7 SUPPLEMENTARY MATERIALS

The source code used to generate all the results in the manuscript are available online at https://github.com/NEU-ABLE-LAB/tidal_grid_open

# 8 AUTHOR CONTRIBUTIONS

Conceptualization, M.K.; methodology, M.K. and J.C; software, A.M, M.K., J.C., F.O. and K.G. ; validation, M.K.; formal analysis, M.K.; investigation, A.M, M.K., J.C., and F.O.; resources, A.M. M.K., F.O. and K.G.; data curation, A.M, M.K., J.C., and F.O.; writing—original draft preparation, A.M. and M.K.; writing—review and editing, M.K. and K.G.; visualization, A.M, M.K., J.C., F.O. and K.G.; supervision, M.K. and K.G.; project administration, M.K.; funding acquisition, M.K. All authors have read and agreed to the published version of the manuscript.

# 9 FUNDING

None.

# 10 ACKNOWLEDGMENTS

The authors A. Marriott, J. Cohen, and F. Ollivierre would like to thank and acknowledge the support provided by the Northeastern University (NU) Young Scholars Program (YSP) run by the NU Center for STEM Education and directed by Claire Duggan.

The author M. Kane would like to thank Joshua Gallaway, Ph.D., assistant professor of chemical engineering at Northeastern University, for many enlightening conversations on modeling battery costs and lifecycles.

This work was completed in part using the Discovery cluster, supported by Northeastern University's Research Computing team.

# 11 CONFLICTS OF INTEREST

The authors declare no conflict of interest.

# 12 REFERENCES

[1] Z. Zhou, M. E. H. Benbouzid, J. F. Charpentier, and F. Scuiller, "Hybrid Diesel/MCT/Battery Electricity Power Supply System for Power Management in Small Islanded Sites: Case Study for the Ouessant French Island," in *Smart Energy Grid Design for Island Countries*, F. M. R. Islam, K. A. Mamun, and M. T. O. Amanullah, Eds. Cham: Springer International Publishing, 2017, pp. 415–445.

[2] X. Hu, C. Zou, C. Zhang, and Y. Li, "Technological Developments in Batteries: A Survey of Principal Roles, Types, and Management Needs - IEEE Journals & Magazine." https://ieeexplore.ieee.org/document/8011541 (accessed Oct. 21, 2020).

[3] D.-I. Stroe, A. Zaharof, and F. Iov, "Power and Energy Management with Battery Storage for a Hybrid Residential PV-Wind System – A Case Study for Denmark," *Energy Procedia*, vol. 155, pp. 464–477, Nov. 2018, doi: 10.1016/j.egypro.2018.11.033.

[4] G. V. B. Kumar, R. K. Sarojini, K. Palanisamy, S. Padmanaban, and J. B. Holm-Nielsen, "Large Scale Renewable Energy Integration: Issues and Solutions," *Energies*, vol. 12, no. 10, Art. no. 10, Jan. 2019, doi: 10.3390/en12101996.

[5] U. Akram, M. Khalid, and S. Shafiq, "An Innovative Hybrid Wind-Solar and Battery-Supercapacitor Microgrid System—Development and Optimization," *IEEE Access*, vol. 5, pp. 25897–25912, 2017, doi: 10.1109/ACCESS.2017.2767618.

[6] M. Uhrig, S. Koenig, M. R. Suriyah, and T. Leibfried, "Lithium-based vs. Vanadium Redox Flow Batteries – A Comparison for Home Storage Systems," *Energy Procedia*, vol. 99, pp. 35–43, Nov. 2016, doi: 10.1016/j.egypro.2016.10.095.

[7] M. Skyllas-Kazacos, G. Kazacos, G. Poon, and H. Verseema, "Recent advances with UNSW vanadium-based redox flow batteries," *Int. J. Energy Res.*, vol. 34, no. 2, pp. 182–189, Feb. 2010, doi: 10.1002/er.1658.

[8] S. Ha and K. G. Gallagher, "Estimating the system price of redox flow batteries for grid storage," *J. Power Sources*, vol. 296, pp. 122–132, Nov. 2015, doi: 10.1016/j.jpowsour.2015.07.004.

[9] L. Ahmadi, M. Fowler, S. B. Young, R. A. Fraser, B. Gaffney, and S. B. Walker, "Energy efficiency of Li-ion battery packs re-used in stationary power applications," *Sustain. Energy Technol. Assess.*, vol. 8, pp. 9–17, Dec. 2014, doi: 10.1016/j.seta.2014.06.006.

[10] M. Astaneh, R. Roshandel, R. Dufo-López, and J. L. Bernal-Agustín, "A novel framework for optimization of size and control strategy of lithium-ion battery based off-grid renewable

energy systems," *Energy Convers. Manag.*, vol. 175, pp. 99–111, Nov. 2018, doi: 10.1016/j.enconman.2018.08.107.

[11] A. Purvins and Sumner, "Optimal management of stationary lithium-ion battery system in electricity distribution grids," *J. Power Sources*, vol. 242, pp. 742–755, Nov. 2013, doi: 10.1016/j.jpowsour.2013.05.097.

[12] J. Vetter *et al.*, "Ageing mechanisms in lithium-ion batteries," *J. Power Sources*, vol. 147, no. 1–2, pp. 269–281, Sep. 2005, doi: 10.1016/j.jpowsour.2005.01.006.

[13] R. Hemmati and H. Saboori, "Emergence of hybrid energy storage systems in renewable energy and transport applications – A review," *Renew. Sustain. Energy Rev.*, vol. 65, pp. 11–23, Nov. 2016, doi: 10.1016/j.rser.2016.06.029.

[14] W. Jing, "Dynamic modelling, analysis and design of smart hybrid energy storage system for off-grid photovoltaic power systems," p. 209, Jan. 2019.

[15] Y. Wang, H. Yu, M. Yong, Y. Huang, F. Zhang, and X. Wang, "Optimal Scheduling of Integrated Energy Systems with Combined Heat and Power Generation, Photovoltaic and Energy Storage Considering Battery Lifetime Loss," *Energies*, vol. 11, no. 7, Art. no. 7, Jul. 2018, doi: 10.3390/en11071676.

[16] G. K. Singh, "Solar power generation by PV (photovoltaic) technology: A review," *Energy*, vol. 53, pp. 1–13, May 2013, doi: 10.1016/j.energy.2013.02.057.

[17] E. Hittinger, T. Wiley, J. Kluza, and J. Whitacre, "Evaluating the value of batteries in microgrid electricity systems using an improved Energy Systems Model," *Energy Convers. Manag.*, vol. 89, pp. 458–472, Jan. 2015, doi: 10.1016/j.enconman.2014.10.011.

[18] E. Kuznetsova, C. Ruiz, Y.-F. Li, and E. Zio, "Analysis of robust optimization for decentralized microgrid energy management under uncertainty," *Int. J. Electr. Power Energy Syst.*, vol. 64, pp. 815–832, Jan. 2015, doi: 10.1016/j.ijepes.2014.07.064.

[19] K. Y. Lau, C. W. Tan, and A. Yatim, "Photovoltaic systems for Malaysian islands: Effects of interest rates, diesel prices and load sizes," *Energy*, vol. 83, pp. 204–216, Apr. 2015, doi: 10.1016/j.energy.2015.02.015.

[20] M. Jafari, A. Botterud, and A. Sakti, "Estimating revenues from offshore wind-storage systems: The importance of advanced battery models," *Appl. Energy*, vol. 276, p. 115417, Oct. 2020, doi: 10.1016/j.apenergy.2020.115417.

[21] O. H. Mohammed, Y. Amirat, and M. Benbouzid, "Particle Swarm Optimization Of a Hybrid Wind/Tidal/PV/Battery Energy System. Application To a Remote Area In Bretagne, France," *Energy Procedia*, vol. 162, pp. 87–96, Apr. 2019, doi: 10.1016/j.egypro.2019.04.010.

[22] K. S. El-Bidairi, H. Duc Nguyen, S. D. G. Jayasinghe, T. S. Mahmoud, and I. Penesis, "A hybrid energy management and battery size optimization for standalone microgrids: A case study for Flinders Island, Australia," *Energy Convers. Manag.*, vol. 175, pp. 192–212, Nov. 2018, doi: 10.1016/j.enconman.2018.08.076.

[23] G. Bekele and Boneya, "Design of a Photovoltaic-Wind Hybrid Power Generation System for Ethiopian Remote Area," *Energy Procedia*, vol. 14, pp. 1760–1765, Jan. 2012, doi: 10.1016/j.egypro.2011.12.1164.

[24] S. Obara, M. Kawai, O. Kawae, and Y. Morizane, "Operational planning of an independent microgrid containing tidal power generators, SOFCs, and photovoltaics," *Appl. Energy*, vol. 102, pp. 1343–1357, Feb. 2013, doi: 10.1016/j.apenergy.2012.07.005.

[25] M. Javidsharifi, T. Niknam, J. Aghaei, and G. Mokryani, "Multi-objective short-term scheduling of a renewable-based microgrid in the presence of tidal resources and storage devices," *Appl. Energy*, vol. 216, pp. 367–381, Apr. 2018, doi: 10.1016/j.apenergy.2017.12.119.

[26] M. Garcia-Sanz, "Control Co-Design: An engineering game changer," *Adv. Control Appl.*, vol. 1, no. 1, p. e18, 2019, doi: https://doi.org/10.1002/adc2.18.

[27] *MATLAB*. Natick, Massachusetts: The MathWorks Inc., 2020.

[28] "PVWatts Calculator." https://pvwatts.nrel.gov/pvwatts.php (accessed Nov. 20, 2020).

[29] "Useful Life | Energy Analysis | NREL." https://www.nrel.gov/analysis/tech-footprint.html (accessed Jan. 25, 2021).

[30] "Frequency of Tides - The Lunar Day - Tides and water levels: NOAA's National Ocean Service Education." https://oceanservice.noaa.gov/education/tutorial_tides/tides05_lunarday.html (accessed Nov. 20, 2020).

[31] "What Causes Tides? | NOAA SciJinks – All About Weather." https://scijinks.gov/tides/ (accessed Nov. 20, 2020).

[32] A. Vazquez and G. Iglesias, "Capital costs in tidal stream energy projects – A spatial approach," *Energy*, vol. 107, pp. 215–226, Jul. 2016, doi: 10.1016/j.energy.2016.03.123.

[33] K. Mongird *et al.*, "Energy Storage Technology and Cost Characterization Report," PNNL-28866, 1573487, Jul. 2019. doi: 10.2172/1573487.

[34] J. D. Milshtein, "Cost-targeted Design of Redox Flow Batteries for Grid Storage," *J Power Sources*, p. 20, 2015.

[35] F. R. Brushett, M. J. Aziz, and K. E. Rodby, "On Lifetime and Cost of Redox-Active Organics for Aqueous Flow Batteries," *ACS Energy Lett.*, vol. 5, no. 3, pp. 879–884, Mar. 2020, doi: 10.1021/acsenergylett.0c00140.

[36] "Block Island Population & Demographics, Median Income - Point2." https://www.point2homes.com/US/Neighborhood/RI/New-Shoreham/Block-Island-Demographics.html (accessed Nov. 05, 2020).

[37] "Frequently Asked Questions (FAQs) - U.S. Energy Information Administration (EIA)." https://www.eia.gov/tools/faqs/faq.php (accessed Nov. 23, 2020).

[38] R. H. Byrd, M. E. Hribar, and J. Nocedal, "An Interior Point Algorithm for Large-Scale Nonlinear Programming," *SIAM J. Optim.*, vol. 9, no. 4, pp. 877–900, Jan. 1999, doi: 10.1137/S1052623497325107.

[39] "Particle Swarm Optimization Algorithm - MATLAB & Simulink." https://www.mathworks.com/help/gads/particle-swarm-optimization-algorithm.html (accessed Jan. 25, 2021).

[40] *NEU-ABLE-LAB/tidal_grid_open*. ABLE Laboratory at Northeastern University, 2021.

[41] "Wholesale U.S. electricity prices were generally lower and less volatile in 2020 than 2019 - Today in Energy - U.S. Energy Information Administration (EIA)." https://www.eia.gov/todayinenergy/detail.php?id=46396 (accessed Jan. 16, 2021).

[42] "State Electricity Profiles - Energy Information Administration." https://www.eia.gov/electricity/state/ (accessed Jan. 16, 2021).

[43] D. Ray, "Lazard's Levelized Cost of Energy Analysis—Version 11.0," p. 22, 2017.